\documentclass[showpacs,amsmath,amssymb,PRB,superscriptaddress,,reprint]{revtex4}
\usepackage{graphicx}
\usepackage{multirow}
\usepackage{textcomp}
\usepackage{color} 

\usepackage[colorlinks,plainpages=false,linkcolor=blue,urlcolor=blue,citecolor=blue,pdfpagemode=UseNone,pdfstartview=FitBH]{hyperref}
\usepackage{latexsym}        
\usepackage{amssymb}
\usepackage{amsmath}
\usepackage{amsfonts}

\begin{document}
 \title{Effect of Ru susbstitution on atomic displacements in the
 layered SmFe$_{1-x}$Ru$_x$AsO$_{0.85}$F$_{0.15}$ superconductor}

\author{A. Iadecola} 
\affiliation{Dipartimento di Fisica, Universit{\'a}
di Roma ``La Sapienza" - P. le Aldo Moro 2, 00185 Roma, Italy}

\author{B. Joseph} 
\affiliation{Dipartimento di Fisica, Universit{\'a}
di Roma ``La Sapienza" - P. le Aldo Moro 2, 00185 Roma, Italy}

\author{L. Simonelli}
\affiliation{European Synchrotron Radiation Facility, BP220, F-38043
Grenoble Cedex, France}

\author{L. Maugeri} 
\affiliation{Dipartimento di Fisica, Universit{\'a}
di Roma ``La Sapienza" - P. le Aldo Moro 2, 00185 Roma, Italy}

\author{M. Fratini}
\affiliation{Dipartimento di Fisica, Universit{\'a} di Roma ``La
Sapienza" - P. le Aldo Moro 2, 00185 Roma, Italy} 

\author{A. Martinelli}
\affiliation{CNR-SPIN and Universit{\'a} di Genova,
via Dodecaneso 33, 16146 Genova, Italy}

\author{A. Palenzona}
\affiliation{CNR-SPIN and Universit{\'a} di Genova,
via Dodecaneso 33, 16146 Genova, Italy}

\author{M. Putti} 
\affiliation{CNR-SPIN and Universit{\'a} di Genova,
via Dodecaneso 33, 16146 Genova, Italy}

\author{N. L. Saini} 
\affiliation{Dipartimento di Fisica, Universit{\'a}
di Roma ``La Sapienza" - P. le Aldo Moro 2, 00185 Roma, Italy}


\begin{abstract}
The effect of Ru substitution on the local structure of layered
SmFe$_{1-x}$Ru$_x$AsO$_{0.85}$F$_{0.15}$ superconductor has been
studied by As $K$- and Sm $L_3$ - edges x-ray-absorption spectroscopy.
The extended x-ray-absorption fine-structure measurements reveal
distinct Fe-As and Ru-As bondlengths in the Ru substituted samples
with the latter being $\sim$0.03 \AA\ longer.  Local disorder
induced by the Ru substitution is mainly confined to the FeAs
layer while the SmO spacer layer sustains a relative order,
consistent with the x-ray-absorption near-edge structure spectra.  The
results suggest that, in addition to the order/disorder in the active
active iron-arsenide layer, its coupling to the rare-earth\textminus oxygen spacer layer needs to be considered for describing the electronic properties of
these layered superconductors.

\bigskip

\noindent Journal reference : {\it Physical Review B } \href{http://link.aps.org/doi/10.1103/PhysRevB.85.214530}{85 (2012) 214530}\\
DOI: \href{http://dx.doi.org/10.1103/PhysRevB.85.214530}{ 10.1103/PhysRevB.85.214530}
\end{abstract}
\pacs{74.70.Xa;74.62.Dh; 61.05.cj; 78.70.Dm}

\maketitle

\newpage

\section{Introduction}
Since the discovery of high\textminus T$_{c}$ superconductivity in doped
LaFeAsO, the iron-based superconductors continue to attract
substantial interest of the condensed-matter community, producing a
large amount of experimental and theoretical
works \cite{rev2,rev3,rev4,rev5,Boeri_11}.  Among these, the $R$FeAsO ($R$ stands for rare-earth), the so-called 1111-type superconductors (with highest
$T_{c}$ of 55 K for the SmFeAsO$_{1-x}$F$_{x}$), are highly studied
materials.  However, interplay of different electronic degrees of
freedom makes it difficult to distinctly identify the role of
different physical parameters governing the fundamental electronic
structure of these superconductors.  One of the key features is the
layered structure with active FeAs layers separated by $R$O spacer
layers.  The fundamental electronic structure, characterized by Fe 3$d$
interacting with the As 4$p$ states \cite{Boeri_11}, is generally
manipulated by controlling the $R$O spacers, e.g. by substitution at
the $R$ site and/or by substitution at the O
site \cite{rev2,rev3,rev4,rev5}.  In this regard, the interlayer
interaction between the active FeAs layers and the spacer layers
is of particular interest, with atomic disorder and local strain being
the key issues.  Earlier we have addressed these issues where
properties of the 1111-system were manipulated by substitution at the
$R$ site, providing important information on the interaction between
the two layers \cite{Iadecola_EPL09,BJoseph_09,Xu_10,RicciSUST_10}.

Recently, several efforts are made to manipulate the effect of
disorder in the active FeAs layers by substitution.  In
particular, isovalent substitution at the Fe
site \cite{Tropeano_PRB10,Sanna_PRL11,Kitagawa_PRB11,Lee_PRB09} has been used to study the effect of disorder on the
superconductivity.  Apart from the effect on the superconductivity,
these studies were also motivated by the possiblity to answer the question
of the symmetry of order parameter in the iron-based
superconductors \cite{symmetry_RPP11,Chubukov_rev}, either in the
spin-fluctuation approach \cite{Mazin_PRL08,Kemper_NJP10,Kuroki_PRB09}
with anion height as a key parameter \cite{Kuroki_PRB09}, or the
orbital fluctuations mediated superconductivity with
iron-phonons \cite{Kontani_PRL10}.  In this respect it is important to
quantify the disorder induced by the isovalent substitution and to
study its implication on the inter-layer correlations and electronic
properties.  Aiming to address these issues we have studied the local
structure of SmFe$_{1-x}$Ru$_x$AsO$_{0.85}$F$_{0.15}$ superconductors
as a function of Ru substitution.  At the optimum F doping, it has
been found that the Ru substitution in the
SmFe$_{1-x}$Ru$_x$AsO$_{0.85}$F$_{0.15}$ strongly affects the $T_{c}$
with a concomitant enhancement of the disorder as seen by transport
measurements \cite{Tropeano_PRB10}.  Interestingly, the short-range
static magnetic order recovers with the Ru substitution, revealed by muon spin
resonance ($\mu$SR) measurements \cite{Sanna_PRL11}.  Here, to
investigate the nature of the local disorder, we have exploited atomic
site selective x-ray-absorption spectra measured at the As $K$- and Sm
$L_3$- edges providing direct information on the FeAs layers and
the SmO spacer layers.  We find that the local disorder induced by Ru
substitution is mainly confined to the FeAs layers, revealed by
the As $K$-edge extended x-ray-absorption fine-structure (EXAFS)
measurements.  On the other hand, x-ray-absorption near-edge structure
(XANES) spectra at the As $K$- edge combined with Sm $L_3$-edge data
indicate reduced disorder in the SmO spacer layer with increasing Ru
substitution.  These results underline the importance of spacer layers
and the interlayer coupling in these layered superconductors.

\section{Experimental details}

Polycrystalline samples of SmFe$_{1-x}$Ru$_x$AsO$_{0.85}$F$_{0.15}$ ($x$
= 0.0, 0.25, and 0.5) were used for the present study.  Details on the
sample preparation and characterization are given elsewhere
\cite{Tropeano_PRB10}.  The superconducting transition temperatures
($T_{c}$) are 51, 14, and 8 K respectively for the samples with
$x$ = 0.0, 0.25, and 0.5.  The As $K$-edge (E = 11868 eV) and Sm
$L_3$-edge (E = 6717 eV) x-ray-absorption measurements were performed
in transmission mode at the beamline BM23 of the European Synchrotron
Facility (ESRF), Grenoble.  A minimum of five scans were acquired on each
samples to ensure high signal to noise ratio and the spectral
reproducibility.  The EXAFS oscillations were extracted from the
absorption spectra using standard procedure \cite{Konings}.  While the
As $K$-edge EXAFS could be obtained up to high $k$-value, the $k$-range of
the Sm $L_3$-edge EXAFS was limited by the Fe $K$-edge absorption jump
at 7112 eV. The XANES spectra were normalized to the atomic absorption
estimated by a linear fit to the data in the EXAFS region after a
pre-edge background subtraction.

\section{Results and discussions}

\begin{figure}
\includegraphics[width=8.0 cm]{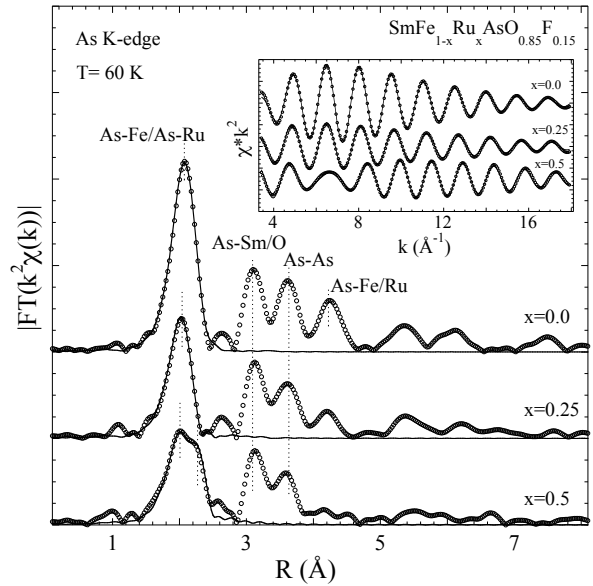}
\caption{\label{fig:1}
Fourier transform (FT) magnitudes of the As $K$-edge EXAFS (weighted by
$k^{2}$) measured on SmFe$_{1-x}$Ru$_x$AsO$_{0.85}$F$_{0.15}$ samples
at 60 K (symbols) with the model fits (solid line) considering the
nearest-neighbours (Fe/Ru shell).  The FTs are performed in
the $k$-range of 3-18\AA$^{-1}$ using a Gaussian window.  The inset
shows experimental filtered EXAFS oscillations (symbols) with the
model fits (solid lines).}
\end{figure}

Figure \ref{fig:1} shows Fourier transform (FT) magnitudes of the As $K$-edge ($k$ range 3 - 18 \AA$^{-1}$) EXAFS oscillations, measured on
SmFe$_{1-x}$Ru$_x$AsO$_{0.85}$F$_{0.15}$ at 60 K, providing partial
atomic distribution around the As atoms.  There are four Fe/Ru near
neighbours of arsenic at a distance $\sim$ 2.4 \AA\ and their
contribution appears as the main peak in the FT at $\sim$ 2 \AA.
The next nearest neighbours of arsenic are Sm ($\sim$ 3.3 \AA) and
O/F ($\sim$ 3.5 \AA) atoms, follwed by the As atoms at $\sim$ 3.9 \AA.
Contributions of these distant shells appear mixed with the multiple
scattering contribution due to Fe/Ru ($\sim$ 4.6 \AA), appearing as FT
peaks in the range $\sim$ 3-5 \AA. The amplitude of the main FT peak
is strongly damped with the Ru substitution.  Compared to this, the FT
peak due to As-Sm/O shows negligible change while the As-As scattering
appears to suffer a small decrease.  In the sample with $x$ = 0.25, the
As-Fe/Ru peak is decreased by almost half and appears as a clear
doublet structure in the sample with $x$ = 0.5.  Also the multiple
scattering peak due to As-Fe/Ru sustains large change, almost
disappearing for the sample with $x$ = 0.5.  These observations suggest
that the atomic disorder introduced by the Ru is confined mainly to
the FeAs layer, with minor influence on the SmO spacer layer.

To quantify the disorder, we have analyzed the first shell EXAFS
containing contribution only due to the As-Fe/Ru bonds, well separated
from other contributions.  In the single-scattering approximation, the
EXAFS amplitude is described by the following general
equation \cite{Konings}:
\begin{equation}
\chi(k)= \sum_{i}\frac{N_{i}S_{0}^{2}}{k R_{i}^{2}}f_{i}(k,R_{i})
e^{-\frac{2R_{i}}{\lambda}} e^{-2k^{2}\sigma_{i}^{2}}
sin[2kR_{i}+\delta_{i}(k)]\nonumber
\end{equation}
where $N_{i}$ is the number of neighbouring atoms at a distance
$R_{i}$ from the photoabsorber.  $S_{0}^{2}$ is the passive electrons
reduction factor, $f_{i}$(k,R$_{i}$) is the backscattering amplitude,
$\lambda$ is the photoelectron mean free path, $\delta_{i}$ is the
phase shift, and $\sigma_{i}^{2}$ is the correlated Debye-Waller
factor (DWF) measuring the mean-square relative displacements (MSRDs)
of the photoabsorber-backscatterer pairs.

The filtered EXAFS oscillations are displayed as the inset of the Fig. \ref{fig:1},
revealing clear damping with Ru substitution.  In the model fits we
have varied the As-Fe/Ru distances and the Debye-Waller factor
($\sigma^{2}$), while all other parameters (photo-electron energy
origin E$_{0}$, the number of near neighbors $N_{i}$ and $S_{0}^{2}$)
were kept fixed in the least squares modelling with structural input
from diffraction studies \cite{Tropeano_PRB10}.  Phase shifts and
amplitude factors were calculated using the {\tiny{FEFF}} code \cite{Feff}.
The number of independent data points for this analysis was 11
(N$_{ind}\sim$(2$\Delta k \Delta$R)/$\pi$, where $\Delta k$ = 15
\AA$^{-1}$ and $\Delta$R = 1.2\AA$ $ are the ranges in $k$ and R space
over which the data are analyzed) for the two (four) parameters fit to
the EXAFS of unsubstituted (substituted) sample.

\begin{table*}[h]
\caption{\label{tab:1}Near neighbour distances and their
$\sigma^{2}$ measured by EXAFS for the
SmFe$_{1-x}$Ru$_x$AsO$_{0.85}$F$_{0.15}$ as a function of Ru
substitution (T = 60 K).  The average uncertaities, determined by
correlation maps, are $\pm$0.006 and $\pm$0.0004 respectively for the
distances and and $\sigma^{2}$ determined by the As $K$-edge.  The
uncertainties for the parameters obtained by the Sm $L_{3}$ edge 
are almost twice those for the As K edge.}
\begin{tabular}{ccccccccccc}\hline
&R$_{Fe-As}$(\AA)&$\sigma_{Fe-As}^{2}$(\AA$^{2}$)&&R$_{Ru-As}$(\AA)&$\sigma_{Ru-As}^{2}$(\AA$^{2}$)&&R$_{Sm-O}$(\AA)&$\sigma_{Sm-O}^{2}$(\AA$^{2}$)\\
\hline
$x$ = 0.0&2.392&0.0030&&-&-&&2.288&0.0054\\
$x$ = 0.25&2.387&0.0032&&2.419&0.0030&&2.285&0.0049\\
$x$ = 0.50&2.390&0.0052&&2.429&0.0025&&2.291&0.0040\\
\hline
\end{tabular}
\end{table*}

The bond distances and the $\sigma^{2}$ (describing mean square
relative displacements) obtained from the above analysis are given in
Table \ref{tab:1}  The Fe-As distance is found to remain constant about
$\sim$2.39 \AA\ for different $x$, however, this distance differs
from the Ru-As distance, measured to be about $\sim$2.42 \AA. The
difference between the two bonds ($\sim$0.03 \AA) in this system is
smaller than that measured ($\sim$0.06 \AA) in the couples of
isostructural compounds RuAs - FeAs and RuAs$_{2}$ -
FeAs$_{2}$ \cite{Heyding_CJC}.  Thus, it appears that the FeAs$_{4}$
(RuAs$_{4}$) blocks are under chemical pressure in the 1111-structure
by the $R$O layers.  On the other hand, the Fe-As bond in these
materials is known to be highly covalent \cite{Iadecola_EPL09}.  In fact, this bond in the 1111 system hardly shows any change even with the changing rare-earth size \cite{Iadecola_EPL09,Martinelli_PRB09}.  Again, the local Fe-As
distance in SmFe$_{1-x}$Ru$_x$AsO$_{0.85}$F$_{0.15}$ with Ru
substitution remains constant, consistent with its highly covalent
nature.  Incidentally, the correponding $\sigma^{2}$ for the Fe-As
bond lengths are similar for samples with $x$ = 0.0 and 0.25, even if
there is a clear increase in $\sigma^{2}$ of this bond for the $x$ =
0.50 sample, with a small decrease of $\sigma^{2}$ for the Ru-As bonds
(Table \ref{tab:1}).  It is known that, while the $T_{c}$ of the
SmFe$_{1-x}$Ru$_x$AsO$_{0.85}$F$_{0.15}$ decreases from 51 to 14 K
with $x$ = 0 to $x$ = 0.25, the residual resistivity increases with a local
maximum around $x$ = 0.25, that has been assigned to impurity scattering
in the system \cite{Tropeano_PRB10}.  Incidentally, the residual
resistivity decreases from $x$ = 0.25 to 0.5, while the $T_{c}$ shows a
smaller change from 14 to 8 K \cite{Tropeano_PRB10}.  Therefore, in the
light of present findings, it is likely that the $x$ = 0.5 sample gets
phase separated unlike the samples with $x \le$0.25 in which the Ru
appear as impurity centers.

\begin{figure}
\includegraphics[width=8.0 cm]{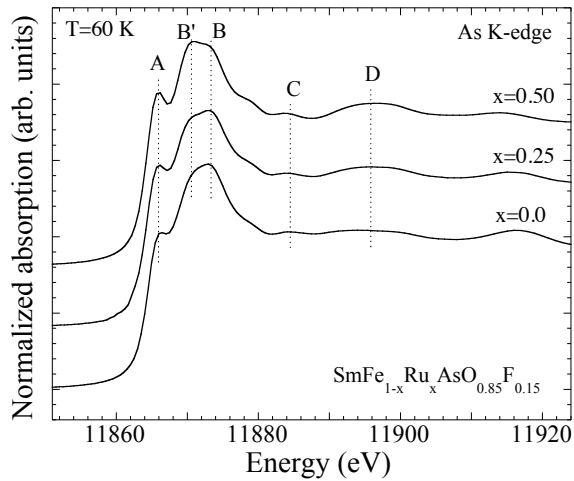}
\caption{\label{fig:2}
Arsenic $K$-edge XANES of SmFe$_{1-x}$Ru$_x$AsO$_{0.85}$F$_{0.15}$ measured
at 60 K. Different near edge features are marked as A, B$^{\prime}$, B, C
and D.}
\end{figure}

Being a probe of higher order atomic correlations, the XANES
measurements provide important information on the local geometry.
Figure 2 shows the As $K$-edge XANES spectra of the
SmFe$_{1-x}$Ru$_x$AsO$_{0.85}$F$_{0.15}$ samples.  Different near edge
features are marked as A, B$^{\prime}$, B, C and D. The ground state
electronic configuration of As atom is [Ar]3$d^{10}$4$s^2$4$p^3$ and the
As $K$-edge spectra probes the transition of core 1$s$ electrons to the
empty $p$ states.  Multiple scattering (MS) calculations of As $K$-edge
XANES features for $R$FeAsO have shown that the absorption feature A
has predominant As 4$p$ character with admixed Fe/Ru $d$ states.
Similarly, the feature B is due to As 4$p$ admixed with Fe/Ru $p$ states.
Also, the distant features C and D appear to have predominantly As 4$p$
character \cite{Xu_10}.  With increased Ru doping the intensity of the
feature A increases, indicating increased unoccupied states of
$p$-symmetry, merely due to the extended Ru 4$d$ states compared to the Fe
3$d$ states, and hence larger mixing with the As 4$p$ states.  MS
calculations have further revealed that the feature B$^{\prime}$ is
correlated with the local geometry of the $R$O spacers.  Indeed this
feature is relatively intense in the As $K$-edge XANES spectrum for
larger $R$ (e.g., LaFeAsO) with respect to the smaller $R$ (e.g.,
SmFeAsO) system.  This is due to higher local disorder of $R$O layer in
the system with smaller $R$ size (i.e., SmO) with respect to the system
with larger $R$ (i.e., LaO) \cite{BJoseph_09,Xu_10}.  The feature
B$^{\prime}$ in the spectra of
SmFe$_{1-x}$Ru$_x$AsO$_{0.85}$F$_{0.15}$ gets intense with increasing
Ru substitution that is an indication of reduced disorder in the SmO
spacer layer.  In this respect, it appears that the effect of the Ru
substitution in the FeAs layer is similar to the one with increasing
rare-earth size \cite{BJoseph_09,Xu_10} (i.e., reduced
disorder in the $R$O spacer).  Therefore, the oxygen order/disorder in
the $R$O spacer layer should have significant role in the properties of
SmFe$_{1-x}$Ru$_x$AsO$_{0.85}$F$_{0.15}$ as well.

\begin{figure}
\includegraphics[width=8.0 cm]{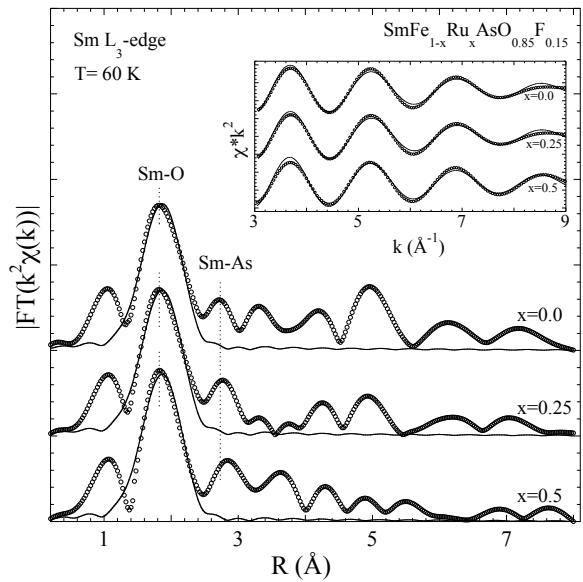}
\caption{\label{fig:3}
Fourier transform (FT) magnitudes of the Sm $L_3$-edge EXAFS (weighted
by $k^{2}$) measured on SmFe$_{1-x}$Ru$_x$AsO$_{0.85}$F$_{0.15}$
($x$=0.0, 0.25, 0.5) samples at 60 K (symbols) with the model fits
(solid line).  The FTs are performed in the $k$-range of
3.0-9.0 \AA$^{-1}$ using a Gaussian window.  The inset shows
experimental filtered EXAFS oscillations with the model fits (solid
lines).}
\end{figure}

To address the question of order/disorder in the spacer layer, we have
directly measured the atomic correlations by Sm $L_{3}$-edge
absorption spectroscopy.  Figure \ref{fig:3} shows FT magnitudes of the Sm
$L_3$-edge ($k$ range 3 - 9.0 \AA$^{-1}$) EXAFS oscillations at 60 K,
providing the atomic distribution around Sm.  For the Sm site, there
are four O near neighbours at a distance $\sim$ 2.3 \AA\ (main peak
at $\sim$ 1.8 \AA).  The next neighbours of Sm are four As atoms at
a distance $\sim$ 3.3 \AA, and Fe atoms at a distance $\sim$ 3.6 \AA.
An apparent shift of the Sm-As FT peak position may be due to
increasing Sm-As distance, consistent with the As $K$-edge EXAFS,
however limited $k$-range of the data does not permit us to make any
further quantification.  Here, we focus only on the Sm-O bonds, the
contribution of which is well separated from the contributions of the
distant shells and hence can be analysed using a single shell model.
The filtered Sm-O EXAFS oscillations are also included in Fig.  \ref{fig:3}
(inset).  Following a similar approach as above, we have kept fixed all
the parameters except the Sm-O distance and the $\sigma^{2}$.  The
number of independent data points for this analysis was about 4 for
the two parameters fit.  Although the k-range is limited, a single
shell analysis with two parameters can still provide useful
information on the near neighbor displacements with a good confidence
level.  The Sm-O distance and its $\sigma^{2}$, determined by the
single shell modeling, are included in the Table \ref{tab:1}.  The results
suggest that the Sm-O distance remains constant within the
experimental uncertainties, however, the $\sigma^{2}$ tend to decrease
with the Ru substitution.  This implies that the local disorder in the
SmO layer is getting reduced, consistent with the conclusions drawn on
the basis of As $K$-edge XANES (Fig.  \ref{fig:2}).

\begin{figure}
\includegraphics[width=8.0 cm]{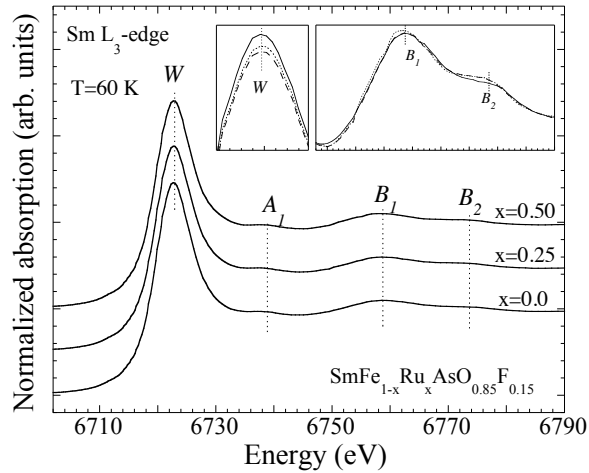}
\caption{\label{fig:4} 
Sm $L_3$-edge XANES spectra of
SmFe$_{1-x}$Ru$_x$AsO$_{0.85}$F$_{0.15}$ ($x$ = 0.0, 0.25, and 0.5).
The whiteline is indicated by W while other near edge features are
marked as A, B$_1$ and B$_2$.  The insets show zoom over the whiteline
(left) and the features A, B$_1$ and B$_2$ (right).  The solid, dashed
and dotted lines in the insets correspond to the spectra of $x$ = 0.0,
0.25, and 0.5 respectively.}
\end{figure}

To have further information, we have analyzed the Sm $L_3$-edge XANES
spectra.  Figure \ref{fig:4} shows normalized Sm $L_3$-edge XANES spectra
measured on SmFe$_{1-x}$Ru$_x$AsO$_{0.85}$F$_{0.15}$ as a function of
Ru substitution.  The spectra show an intense peak, the characteristic
white line (W) of Sm$^{3+}$ due to 2p$_{3/2}\to$5$\epsilon$d
transition.  The other near edge features are denoted by A$_{1}$,
B$_{1}$ and B$_{2}$, appearing around 15 eV, 35 eV, 50 eV above the
white line.  Compared to the As K-edge XANES the changes in the Sm
$L_{3}$-edge spectra are much smaller, confirming once again that the
main effect of the Ru substitution is confined to the FeAs layer.
The features B$_{1}$ and B$_{2}$ are continuum resonance
peaks \cite{BJoseph_09} due to scatterings with As at $\sim$ 3.3
\AA\ and O at $\sim$ 2.3 \AA.  MS calculations of RE $L_{3}$-edge
in the 1111 system have shown that the feature A$_{1}$ is sensitive to
the order/disorder in the SmO plane \cite{Xu_10}.  Although
small, there are few apparent changes in the XANES features.  While
the B$_{1}$ shifts towards lower energy, the B$_{2}$ appears to gain
some intensity with Ru substitution (inset in Fig.  \ref{fig:4}).
The shift of B$_{1}$ is due to increased Sm-As distance, consistent
with the EXAFS. On the other hand, the increased intensity of the
B$_{2}$ indicates reduced disorder in the SmO sublattice with Ru
substitution, again consistent with the EXAFS data.

In summary, we have studied the local structure of the superconducting
SmFe$_{1-x}$Ru$_x$AsO$_{0.85}$F$_{0.15}$ system with variable Ru by
x-ray absorption measurements.  The EXAFS and XANES spectra obtained
at the As $K$- and Sm $L_{3}$-edges have permitted to obtain information
on the local structure of the active FeAs layers and the SmO
spacer layers.  The EXAFS data reveal distinct Fe-As and Ru-As
bondlengths and the local disorder being confined in the FeAs
layer, while the SmO spacer layer sustaining a local order.
Therefore, the effect of Ru substitution in the FeAs layer appears
similar to what has been found in the rare-earth substituted $R$FeAsO
with different rare-earth size \cite{Iadecola_EPL09,BJoseph_09,Xu_10}.  
It was found that the coupling between the two layers gets weaker for bigger
size of rare-earth while the FeAs layers getting thinner, similar
to the increasing Ru substitution at the Fe site seen in this work.

The decoupling of the two layers means that the two sublattices are
independent and the active FeAs layer looses screening from the spacer
$R$O layer due to smaller interlayer coupling in the Ru substituted
system.  On the other hand, the FeAs layers get thinner,
consistent with diffraction \cite{Tropeano_PRB10}, and electronically
the system acts as a 11-type superconductor (i.e., a binary FeSe),
albeit with smaller pnictogen height from the Fe-plane, i.e., more
more three-dimensional character of the band structure (larger k$_{z}$-dispersion as seen
by angle resolved photoemission \cite{Xu_Condmat12}) near the Fermi
level that is mainly derived by the Fe 3$d$ admixed As 4$p$ orbitals.
Smaller interlayer coupling also mean that the substituted Ru atoms in
the FeAs layers acting as localized disorder due to poorer
screening and/or there is a phase separation similar to the one
appears in the ternary FeSe$_{1-x}$Te$_{x}$ systems on Te
substitution \cite{BJoseph_PRB10}.  Here, It seems that the impurity
scattering due to the atomic disorder being partly responsible for the
$T_c$ suppression, also evident from the residual resistivity
behaviour with the Ru concentration \cite{Tropeano_PRB10}.  Since
residual resistivity decreases as well the interlayer coupling with
further Ru concentration ($x \ge$0.25), it is likely that the static
disorder prevails while the system gets phase separated with reentrant
local magnetic order \cite{Sanna_PRL11}.  In the present case, the
$T_c$ decreases from 51 to 14 K with smaller amounts of Ru substitution
($x$ = 0.0 to 0.25), while the $T_c$ change is smaller, from 14 to 8 K, with larger amounts of Ru substitution (for $x$ = 0.25 to 0.5).
Therefore different mechanisms appears to be active for the T$_c$
suppression with different concentration ranges for the isoelectronic substitution.  In conclusion, the present results demonstrate importance of interlayer atomic correlations for describing the electronic properties of the layered 1111-type superconductors.  Having direct implication on spin/orbital
fluctuation theories \cite{symmetry_RPP11,Chubukov_rev,Mazin_PRL08,Kemper_NJP10,Kuroki_PRB09,Kontani_PRL10}, these topological aspects need proper consideration for a realistic
description of the superconductivity in these materials.

\begin{acknowledgments}
Experimental support by the beamline staff at the ESRF is kindly
acknowledged. The work at Genova is partially supported by FP7 SUPER-IRON Project (No.283204).
\end{acknowledgments}


\begin{thebibliography}{100}


\bibitem{rev2} D. C. Johnston, 
\href{http://dx.doi.org/10.1080/00018732.2010.513480}{
Advances in Physics 59, 803 (2010)}. 

\bibitem{rev3}G. R. Stewart,  
\href{http://dx.doi.org/10.1103/RevModPhys.83.1589}{
Reviews of Modern Physics 83, 1589 (2011)}.

\bibitem{rev4}P. C. Canfield and S. L. Bud'ko, 
\href{http://dx.doi.org/10.1146/annurev-conmatphys-070909-104041}{
Annual Review of Condensed Matter Physics  1, 27 (2010)}.

\bibitem{rev5}H. H. Wen and S. Li, 
\href{http://dx.doi.org/10.1146/annurev-conmatphys-062910-140518}{
Annual Review of Condensed Matter Physics  2, 121 (2011)}.

\bibitem{Boeri_11}O.K. Andersen and L. Boeri, 
\href{http://dx.doi.org/10.1002/andp.201000149}{
Annalen der Physik 523, 8 (2011)}.

\bibitem{Iadecola_EPL09} A. Iadecola, S. Agrestini, M. Filippi, L.
Simonelli, M. Fratini, B. Joseph, D. Mahajan and N. L. Saini, 
\href{http://dx.doi.org/10.1209/0295-5075/87/26005}{
Europhysics Letters 87, 26005 (2009)}.

\bibitem{BJoseph_09} B. Joseph, A. Iadecola, M. Fratini, A. Bianconi,
A. Marcelli and N.L. Saini, 
\href{http://dx.doi.org/10.1088/0953-8984/21/43/432201}{
Journal of Physics: Condensed Matter 21,
432201 (2009)}; W. Xu, A. Marcelli, B. Joseph, A. Iadecola, W. S. Chu,
D. Di Gioacchino, A. Bianconi, Z. Y. Wu, and N. L. Saini, 
\href{http://dx.doi.org/10.1088/0953-8984/22/12/125701}{
Journal of Physics: Condensed Matter 22, 125701 (2010)}.

\bibitem{Xu_10} W. Xu, B. Joseph, A. Iadecola, A. Marcelli, W.S. Chu,
D. Di Gioacchino, A. Bianconi, Z. Y. Wu and N. L. Saini, 
\href{http://dx.doi.org/10.1209/0295-5075/90/57001}{
Europhysics Letters 90, 57001 (2010)}.

\bibitem{RicciSUST_10}A. Ricci, B. Joseph, N. Poccia, W. Xu, D. Chen,
W. S. Chu, Z. Y. Wu, A. Marcelli, N. L. Saini, and A. Bianconi,
\href{http://dx.doi.org/10.1088/0953-2048/23/5/052003}{
Superconductor Science and Technology 23, 052003 (2010)}.

\bibitem{Tropeano_PRB10} M. Tropeano, M. R. Cimberle, C. Ferdeghini,
G. Lamura, A. Martinelli, A. Palenzona, I. Pallecchi, A. Sala, I.
Sheikin, F. Bernardini, M. Monni, S. Massidda, and M. Putti, 
\href{http://dx.doi.org/10.1103/PhysRevB.81.184504}{ 
Physical
Review B 81, 184504 (2010)}; I. Pallecchi, F. Bernardini, M. Tropeano,
A. Palenzona, A. Martinelli, C. Ferdeghini, M. Vignolo, S. Massidda,
M. Putti, 
\href{http://dx.doi.org/10.1103/PhysRevB.84.134524}{
Physical Review B 84, 134524 (2011)}.

\bibitem{Sanna_PRL11} S. Sanna, P. Carretta, P. Bonfa, G. Prando, G.
Allodi, R. De Renzi, T. Shiroka, G. Lamura, A. Martinelli, and M.
Putti, \href{http://dx.doi.org/10.1103/PhysRevLett.107.227003}{ 
Physical  Review  Letters  107, 227003 (2011)}.

\bibitem{Kitagawa_PRB11} S. Kitagawa, Y. Nakai, T. Iye, K. Ishida, Y.
F. Guo, Y. G. Shi, K. Yamaura, and E. Takayama-Muromachi, 
\href{http://dx.doi.org/10.1103/PhysRevB.83.180501}{
Physical Review B
83, 180501 (2011)}.

\bibitem{Lee_PRB09} S. C. Lee, A. Kawabata, T. Moyoshi, Y. Kobayashi,
and M. Sato, \href{http://dx.doi.org/10.1143/JPSJ.78.043703}{
Journal of the Physical Society of Japan 78, 043703
(2009)}; Y. Kobayashi, A. Kawabata, S. C. Lee, T. Moyoshi, and M. Sato,
\href{http://dx.doi.org/10.1143/JPSJ.78.073704}{
Journal of the Physical Society of Japan 78, 073704 (2009)}; 
M. Sato, Y. Kobayashi, S. C. Lee, H. Takahashi, E. Satomi, and Y. Miura,
\href{http://dx.doi.org/10.1143/JPSJ.79.014710}{
Journal of the Physical Society of Japan 79, 014710 (2010)}.

\bibitem{symmetry_RPP11} J. Hirschfeld, M. M. Korshunov and I. I.
Mazin, \href{http://dx.doi.org/10.1088/0034-4885/74/12/124508}{  
Report on Progress in Physics  74 124508 (2011)}.

\bibitem{Chubukov_rev}A. Chubukov, 
\href{http://dx.doi.org/10.1146/annurev-conmatphys-020911-125055}{ Annual Review of Condensed Matter Physics 3,57 (2012)}.

\bibitem{Mazin_PRL08}I. I. Mazin, D. J. Singh, M. D. Johannes, and M.
H. Du, 
\href{http://dx.doi.org/10.1103/PhysRevLett.101.057003}{
Physical  Review  Letters  101, 057003 (2008)}.

\bibitem{Kemper_NJP10}A.F. Kemper, T.A. Maier, S. Graser,H-P. Cheng,
P.J. Hirschfeld and D.J. Scalapino,
\href{http://dx.doi.org/10.1088/1367-2630/12/7/073030}{
New Journal of Physics  12, 073030 (2010)}.

\bibitem{Kuroki_PRB09}K. Kuroki, H. Usui, S. Onari, R. Arita, and H.
Aoki, 
\href{http://dx.doi.org/10.1103/PhysRevB.79.224511}{ 
Physical Review B 79, 224511 (2009)}.

\bibitem{Kontani_PRL10}H. Kontani and S. Onari, 
\href{http://dx.doi.org/10.1103/PhysRevLett.104.157001}{
Physical Review Letters
104, 157001 (2010)}.

\bibitem{Konings}X-ray Absorption: Principles, Applications,
Techniques of EXAFS, SEXAFS, XANES, edited by R. Prins and D. C.
Koningsberger (Wiley, New York, 1988).

\bibitem{Feff}J. Mustre de Leon, J. J. Rehr, S. I. Zabinsky, and R. C.
Albers, \href{http://dx.doi.org/10.1103/PhysRevB.44.4146}{ 
Physical Review B 44, 4146 (1991)};
J. J. Rehr and R. C. Albers, 
\href{http://dx.doi.org/10.1103/RevModPhys.72.621}{
Reviews of Modern Physics 72, 621 (2000)}.

\bibitem{Heyding_CJC} R.D. Heyding, L.D. Calvert, 
\href{http://dx.doi.org/10.1139/v57-065}{
Canadian Journal of Chemistry  35,
449 (1957)}; R.D. Heyding, L.D. Calvert, 
\href{http://dx.doi.org/10.1139/v61-118}{
Canadian Journal of Chemistry  39, 955 (1961)}.

\bibitem{Martinelli_PRB09}A. Martinelli, A. Palenzona, M. Tropeano, C. Ferdeghini, M. R. Cimberle, and C. Ritter,
\href{http://dx.doi.org/10.1103/PhysRevB.80.214106}{
Physical Review B 80, 214106 (2009)}.

\bibitem{Xu_Condmat12}N. Xu, T. Qian, P. Richard, Y.-B. Shi, X.-P.
Wang, P. Zhang, Y.-B. Huang, Y.-M. Xu, H. Miao, G. Xu, G.-F.
Xuan, W.-H. Jiao, Z.-A. Xu, G.-H. Cao, H. Ding,	
\href{http://arxiv.org/abs/arXiv:1203.4699}{
arXiv:1203.4699v1} (unpublished).

\bibitem{BJoseph_PRB10}B. Joseph, A. Iadecola, A. Puri, L. Simonelli,
Y. Mizuguchi, Y. Takano, and N. L. Saini, 
\href{http://dx.doi.org/10.1103/PhysRevB.82.020502}{ Physical Review B 
82, 020502 (2010)}.

\end{thebibliography}
\end{document}